\documentclass[conference]{IEEEtran}

\usepackage[pdftex,colorlinks=true,hyperindex,backref,hypertexnames=false]{hyperref}
\newtheorem{thm}{Theorem}[section]
\newtheorem{definition}[thm]{Definition}

\hypersetup{%
pdftitle={Antifragility = Elasticity + Resilience + Machine Learning},
pdfauthor={Vincenzo De Florio},
pdfsubject={Characterization of real-time and open-system behaviours},
pdfkeywords={Modeling, Real-time systems, Open systems, Homomorphism, Behavior, Resilience},
bookmarksopen=true,
bookmarksnumbered=true,
pdfstartview={FitH},
urlcolor=blue,
}%

\usepackage{amssymb}

\usepackage{cite}

\begin{document}
\title{Antifragility $=$ \\Elasticity $+$ Resilience $+$ Machine Learning\\
{\large Models and Algorithms for Open System Fidelity}}
%Elasticity and Resilience
%Algebraic/Behavioural Model\\of Real-timeliness and Synchrony}

%If you're using runningheads you can add an abreviated title for the running head on odd pages using the following
%\titlerunning{abreviated title goes here}
%and an alternative title for the table of contents:
%\toctitle{table of contents title}

%\subtitle{Subtitle Goes Here}

%For a single author
\author{%
\IEEEauthorblockN{Vincenzo De Florio}
\IEEEauthorblockA{PATS research group,
University of Antwerp \& iMinds Research Institute\\
Middelheimlaan 1, 2020 Antwerpen, Belgium\\
Email: \href{mailto:vincenzo.deflorio@uantwerpen.ua}{\nolinkurl{vincenzo.deflorio@uantwerpen.be}}}}

%\author{Vincenzo De Florio}

%For multiple authors:
%\author{First Author Name\inst{1} \and Second Author Name\inst{2}}

%If using runnningheads you can abbreviate the author name on even pages:
%\authorrunning{abbreviated author name}
%and you can change the author name in the table of contents
%\tocauthor{enhanced author name}

%For a single institute

% If authors are from different institutes 
%\institute{First Institute Name \email{email address} \and Second Institute Name\thanks{Thank you to...} \email{email address}}

%to remove your email just remove '\email{email address}'
% you can also remove the thanks footnote by removing '\thanks{Thank you to...}'

\def\C{\hbox{$\mathcal{C}$}}
\def\U{\hbox{$\mathcal{U}$}}
\newcommand*\BitOr{\mathrel{|}}
\def\eqref#1{\hbox{(\ref{#1})}}
\maketitle

\begin{abstract}
	We introduce a model of the fidelity of open systems---fidelity being interpreted here
	as the compliance between corresponding figures of interest in two separate but communicating
	domains. A special case of fidelity is given by real-timeliness and synchrony, in which
	the figure of interest is the physical and the system's notion of time.
	Our model covers two orthogonal aspects of fidelity, the first one focusing on a system's
	steady state and the second one capturing that system's dynamic and behavioural characteristics.
	We discuss how the two aspects correspond respectively to elasticity and resilience and we
	highlight each aspect's qualities and limitations. Finally we sketch the elements of a new model
	coupling both of the first model's aspects and complementing them with machine learning.
	Finally, a conjecture is put forward that the new model may represent a first step
	towards compositional criteria for antifragile systems.
\end{abstract}

%%%%%%%%%%%%%%%%%%%%%%%%%%%%%%%%%
\section{Introduction}\label{s:I}
%%%%%%%%%%%%%%%%%%%%%%%%%%%%%%%%%
As well-known, open systems are those that
continuously communicate and ``interact with other systems outside of themselves''~\cite{Hey98}.
Modern electronic devices~\cite{MuCa11} and cyber-physical systems~\cite{Lee:EECS-2008-8}
are typical examples of open systems that
more and more are being deployed in different shapes and ``things'' around us. Advanced
communication capabilities pave the way towards collective organisation of open systems able to
enact complex collective strategies~\cite{AF83} and self-organise into societies~\cite{Zhu10},
communities~\cite{SDGB10b,DeBl10}, networks~\cite{Latour06}, and organisations~\cite{DF13c}.

One of the most salient aspects of open systems---as well as
a key factor in the emergence of their quality---is given by the compliance
between physical figures of interest and their internal representations.
We call this property as fidelity. A high fidelity makes it possible
to build ``internal'' models of ``external'' conditions, which in turn
can be used to improve important design goals---including
performance and resilience. Conversely, low fidelity
results in unsatisfactory models of the ``world'' and the ``self''---an argument
already put forward in Plato's Cave.

As an example, real-time systems are open systems that mainly
focus on a single figure---physical time. Such figure is ``reified''
as cybertime---an internal representation of physical time. Intuitively,
the more accurately the internal representation reflects the property
of a corresponding physical dimension, the higher will be the quality
exhibited by such class of open systems.

In what follows we consider the more general case of $n$-open systems---systems
that is that interact with environments represented through $n$ context figures.
This means that, through some sensory system and some sampling and conversion
algorithms, each of these $n$ context figures is reified in the form
of an internal variable reflecting the state of the corresponding figure.
These ``reflective variables''~\cite{DB07a} are the computational equivalent
of the biological concept of \emph{qualia\/} (pl. \emph{quale})~\cite{Kanai12}
and represent an open system's primary interface to their domains of intervention (typically,
the physical world.)

This work introduces two models for the fidelity of $n$-open systems.
Each of those models provides a different view
to an $n$-open system's nature and characteristics.

The first model is presented
in Sect.~\ref{s:AM} and mainly focuses on \emph{elasticity\/}
support to fidelity.
Quality is reached through simple schemes with
as limited as possible an overhead and as low as possible
an impact on functional design goals. Resource scheduling, redundancy,
and diversity are mostly applied through worst-case analyses
and at design-time, possibly with simple switching
among Pareto-optimal strategies during the run-time~\cite{MuCa11}.
As mentioned above, the key strategy in this case is elasticity: unfavourable
changes and faults are meant to be masked out and counterbalanced
by provisions that do not require intensive system reconfigurations.
This model considers the system and its intended deployment environments
as known and stable entities (cf. synchronous system model~\cite{DeFl09})
and identifies a snapshot of the system in its intended (viz., ``normal'')
operational conditions.

Conversely, our second model---introduced in Sect.~\ref{s:BM}---is behavioural
and focuses on \emph{resilience\/} support to fidelity. 
Systems and their environments are regarded as \emph{dynamic systems\/}
whose features are naturally drifting in time
(as it is the case, e.g., in the timed-asynchronous system model~\cite{CrFe99}).
Corresponding variations in the operational conditions within and without the system boundaries may be tolerated through
different strategies, and this model
focuses on the quality of the behaviours that a system may employ,
during the run-time, in order to guarantee its fidelity despite those variations.

A discussion is then elaborated in Sect.~\ref{s:D}. Positive and negative aspects of both models are highlighted.
Then it is shown how
the two models may co-exist by distinguishing between normal
and critical conditions. A general scheme for context-conscious switching between elasticity and resilience strategies
is proposed. Said scheme also incorporates a machine learning step such that the system may
acquire some form of ``wisdom'' as a by-product of past history. 
A conjecture is put forward that the general scheme may represent a first preliminary step towards
the engineering of \emph{antifragile\/} systems~\cite{Taleb12}, namely systems not merely able to tolerate
adverse conditions, but rather able to strengthen in the process their ability to do so.

Section~\ref{s:C} finally concludes with a view to our future work.

%design issue (immutable conditions pre-defined at design time in \ref{s:AM}
%here assumption is: it won't change
%
%run-time issues: here, system will drift. Run-time behavior is needed to counterbalance. . .

%%%%%%%%%%%%%%%%%%%%%%%%%%%%%%%%%%%%%
\section{Algebraic Model}\label{s:AM}
%%%%%%%%%%%%%%%%%%%%%%%%%%%%%%%%%%%%%
Here we introduce our first model of fidelity. First the main objects of our treatise are presented
in Sect.~\ref{s:AM:FE}. Section~\ref{s:AM:M} follows and introduces
our Algebraic model based on those objects.

%-----------------------------------------%
\subsection{Formal entities}\label{s:AM:FE}
%-----------------------------------------%

As mentioned in Sect.~\ref{s:I}, open systems are those computer systems that interact with
a domain they are immersed in. A prerequisite to the quality of this interaction is the perception~\cite{DF13b}
of a domain- and application-specific number of figures, say $0<n\in\mathbb{N}$. The quality of the perception services
is the cornerstone to fidelity~\cite{DF12a}, the latter taking the shape of, e.g., optimal performance, strong guarantees
of real-timeliness and safety, high quality of experience, and so forth.

Perception in $n$-open systems is
modelled in what follows as a set of functions $q_i, 0<i\le n$, defined between pairs of Algebraic structures, $\U_i$ and $\C_i, 0<i\le n$,
respectively representing sets of physical properties of interest (called in what follows as ``raw facts'')
and sets of their corresponding computer-based operational representations---their ``reflective variables'', or ``quale''.
``Reflective maps'' is the term we shall use to refer to the above functions.

Let us refer as \U{} and \C{} respectively as to any of the $\U_i$ and $\C_i$, and as $q$ to any 
of the reflective maps.
Thus, if $u\in\U$ is, e.g., the amount of light emitted by a light-bulb, 
then $q(u)$ may for instance be a floating point number stored in some memory cells and quantifying the light currently emitted
by the bulb as perceived by some sensor and as represented by some sampling and conversion algorithm.
A reflective map takes the following general form:
\begin{equation}
q : \U \to \C \label{eq:q}
\end{equation}
and obeys the following Condition:
\begin{equation}
	\forall u_1, u_2\in\U: q(u_1 + u_2) = q(u_1) + q(u_2) + \Delta.\label{eq:cond}
\end{equation}

Here overloaded operator ``$+$'' represents two different operations:
\begin{itemize}
	\item In \U, as in the expression on the left of the equal sign, operator ``$+$''
		is the property resulting from the composition of two congruent physical properties.
		As an example, this may be the amount of light produced by turning on two light-bulbs in a room,
		say light-bulb $l_1$ and light-bulb $l_2$. (Note that the amount of light actually perceived
		by some entity in that room will depend on the relative positions of the light-bulbs and
		the perceiver as well as on the presence of obstructing objects in the room, and other factors.)
	\item In \C, as in the expression on the right of the equal sign, 
		operator ``$+$'' is the algorithm that produces a valid operational representation of
		some property by adding any two other valid representations of the same property.
		In the above example, the operator computes the qualia corresponding to the sum
		of the quale representing the light emitted by $l_1$ with that of $l_2$.
\end{itemize}

Let $\Delta$ be a variable representing any $\Delta_i, 0<i\le n$, in turn defined as
the ``preservation distance'' of reflective map $q_i$, meaning
that $q_i$ would preserve operation ``$+$'' were it not for the extra $\Delta_i$.
Quantity $\Delta$ thus represents an error value that depends on the nature of the involved
properties; their operational representations; and ``the environment'', the latter being
modelled as
a set of context figures representing the hardware and software platforms; the operational conditions;
the user behaviours; and other factors. Environmental
conditions shall be cumulatively represented in what follows as vector $\vec{e}$.

%------------------------------%
\subsection{Model}\label{s:AM:M}
%------------------------------%

Function~\eqref{eq:q} and Condition~\eqref{eq:cond} may be used to characterise concisely the fidelity of
an $n$-open system,
namely how coherent, consistent, and robust is the reflection in $\C_i$ of the physical properties of $\U_i, 0<i\le n$.
Our exemplary focus
in the rest of this section will be that of real-time systems (namely 1-open systems whose
domains of interests are cybertime\footnote{We shall refer in what follows to any artificial concept of time,
	as manifested for instance by the amount of clock ticks elapsed between any two computer-related
	events, as to ``cybertime''.}
and physical time) though this will not affect the generality of our treatise.
In the rest of this section $\C$ will refer to cybertime and $\U$ to physical time.
The corresponding reflective map shall be simply referred to as $q$ while $\Delta$ shall be $q$'s preservation distance.
%though a similar treatise may pertain
%to other properties of interest.

The formal cornerstone of our model is given by the concept of
isomorphism---a bijective map between two Algebraic structures characterised by
a property of operation preservation. As well-known, a function such as reflective map $q$
is a isomorphism if it is bijective
and if the preservation distance $\Delta$ is equal to zero.
In this case the two domains, \C{} and \U, are in perfect correspondence: any action (or composition thereof) occurring
in either of the two structures can be associated with an equivalent action (resp. composition) in the other one.
In the domain of time, this translates in perfect equivalence between the physical and
artificial concept of time---between cybertime and physical time that is.
Different interpretations depend on the domain of reference. As an example, in the domain of safety, the
above correspondence may mean that the consequence of \C-actions in terms of events taking place in \U{}
be always measurable and controllable---and vice-versa.

Obviously the above flawless correspondence only characterises a hypothetically
perfect computer system able to sustain its operation in perfect synchrony
with the physical entities it interacts with---whatever
the environmental conditions may be.
The practical purpose of considering such a system is that, like Boulding's transcendental systems~\cite{Bou56} or
Leibniz's Monads and their perfect power-of-representation~\cite{leibniz2006shorter}, it is
a \emph{reference point}.
By identifying specific differences with respect to said reference point 
we can categorise and partition existing families of systems
and behaviours as per the following definitions.

\begin{definition}[Hard real-time system]\label{d:hrt}
	A hard real-time system is the best real-life approximation of a perfect real-time system.
	Though different from zero, its preservation distance (the $\Delta$ function, representing in this case
	the system's ``tardiness'')
	has a bound range (limited by an upper threshold)
	equal to a ``small'' interval (drifts and threshold are, e.g., one order of magnitude smaller than the
	reference time unit for the exercised service).
%\footnote{For instance, in the case of a service
		%for electrical substation automation~\cite{PDP2001} with cycles of period 10ms, 
		%tolerable drifts and thresholds should be in the range of the hundreds of milliseconds.}).
	A hard real-time system is typically guarded, meaning that the system self-checks
	its preservation distance.
\end{definition}

In what follows we shall call as ``$t$-hard real-time system'' a system that
matches the conditions in Definition~\ref{d:hrt} with threshold $t$.

\begin{definition}[Soft real-time system]\label{d:srt}
	A system is said to be ``soft real-time'' if its preservation distance $\Delta$ is
	statistically bound. As in hard real-time systems, a threshold characterises the $\Delta$
	error function, but that threshold is an average value, namely there is no hard guarantee,
	as it was the case for hard real-time systems, that the error will \emph{never\/} be overcome.
	Both threshold and its standard deviation are ``small''.
	As hard real-time systems, also soft real-time systems are typically guarded---viz.,
	they self-check their tardiness (preservation distance.)
\end{definition}

In what follows we shall call a ``$(t,\sigma)$-soft real-time system'' a system that
matches the conditions in Definition~\ref{d:srt} with average threshold $t$
and standard deviation $\sigma$.

\begin{definition}[Best-effort real-time system]\label{d:bert}
	A system is said to be ``best-effort'' if care and experience have been put to use,
	up to a certain degree\footnote{A trade-off between design quality,
				usability, time-to-market, costs and other factors
				typically affects and limits the employed care.},
	in order to design and craft that system; said care and experience, to the best of
	the current knowledge and practise,
	should allow the $\Delta$ values experienced by the users to be considered as
	``acceptable'', meaning that deviations from the expected behaviours are such that
	the largest possible user audience shall not be discouraged from making use of the system. 
	Internet-based teleconferencing systems are examples of
	systems in this category.
	Unlike hard and soft real-time systems, best effort systems do not monitor
	the drifting of 
	their $\Delta$\footnote{In some cases monitoring data are
		gathered from the users. As an example, the users of the
		Skype teleconferencing
		system are typically asked to provide an assessment of the
		the quality of their experience after using the service.
		This provides the Skype administrators with statistical data
		regarding the $\Delta$'s experienced by their users.}.
\end{definition}

It is important to highlight once more how function $\Delta$ is also a function of $\vec{e}$---the environmental
conditions.
As mentioned already, the above conditions include those pertaining to the characteristics
and the current state of the deployment platform.
As a consequence of this dependency,
special care is required to verify that the system's deployment and run-time hypotheses
\emph{will stay\/} valid
over time. Sect.~\ref{s:BM} specifically covers this aspect.
Assumption failure tolerance~\cite{De10} may be used to detect and treat
deployment and run-time assumption mismatches.

\begin{definition}[Non-real-time system]\label{d:nrt}
A non real-time system is one that is employed, deployed, and executed, with no concern
and no awareness of the drifting of function $\Delta$. With respect to time, the system
is context-agnostic and is meant to be used ``as is''---without any operational or quality guarantee.
\end{definition}

Definitions~\ref{d:hrt}--\ref{d:nrt} can be used to partition systems into
disjoint blocks (or equivalence classes). Said classes may be regarded as ``contracts''
that the systems need to fulfil in order to comply to their (real-timeliness)
specifications.

\begin{definition}[System identity]\label{d:id}
	We define as system real-time identity (in general, its system identity)
	the equivalence class a (real-time) system belongs to.
\end{definition}

We now discuss a second 
and complementary aspect---system behaviour and its effect
on the correspondence between \C{} and \U.

%it is unmovable, no ``motion'', no active or puposeful behaviour.
%unmoved mover
%The just mentioned hypothetical 
%No drifting 
%isomorphism, or bijective homomorphism.

%%%%%%%%%%%%%%%%%%%%%%%%%%%%%%%%%%%%%%%
\section{Behavioural Model}\label{s:BM}
%%%%%%%%%%%%%%%%%%%%%%%%%%%%%%%%%%%%%%%
As already hinted in Sect.~\ref{s:I}, our Algebraic model and its
Definitions~\ref{d:hrt}--\ref{d:nrt} 
do not cover an important aspect of open systems~\cite{Hey98},
namely the fact that, in real-life, the extent and the rate of the
environmental changes may (as a matter of fact, \emph{shall})
produce a sensible effect on $\vec{e}$ (and thus, on $\Delta$)
even when the system has been designed with the utmost care.
In this case the system may experience an ``identity failure'', namely
a drift\footnote{The problem of system identity drift going undetected
	is one that may produce serious consequences---especially in the
	case of safety-critical computer systems. This is efficaciously
	expressed by quoting Bill Strauss: ``A plane is designed to
	the right specs, but nobody goes back and checks
	if it is still robust''~\cite{p3air}.}
that produces a loss of the system identity (cf. Def.~\ref{d:id}). 

In order to capture a system's ability to detect, mask, tolerate, or anticipate
identity failures,
that system needs to enact a number of resilient behaviours~\cite{DF13a,DF13b}.
In what follows we first briefly introduce
the concepts of resilience and behaviours (respectively
in Sect.~\ref{s:BM:R} and Sect.~\ref{s:BM:B}) and then discuss in Sect.~\ref{s:BM:M}
how resilient behaviours
constitute a second ``parameter'' with which one may characterise
salient aspects of the fidelity of open systems.

%-----------------------------------%
\subsection{Resilience}\label{s:BM:R}
%-----------------------------------%
Resilience is a system's ability
to retain certain characteristics of interest throughout changes
affecting itself and its environments.
By referring to Sect.~\ref{s:AM:M} and in particular to Def.~\ref{d:id},
resilience may be defined as \emph{robust system identity persistence}, namely
a system's ``ability to pursue completion (that is,
one's optimal behaviour) by continuously re-adjusting oneself''~\cite{DF13b}.
Resilience closely corresponds to the
Aristotelian concept of entelechy~\cite{Sachs,DeAnimaLawsonTancred},
namely ``exercising activity in order to guarantee
one's identity'', or to ``comply to one's `definition'.''

As suggested in~\cite{DF13a}, resilience calls for (at least) the following three abilities:
\begin{itemize}
	\item Perception, namely the ability
		to become timely aware of some portion of the raw facts in the environment
		(both within and without the system boundaries).
	\item Awareness, which ``defines how
		the reflected [raw facts] are accrued, put in relation with past perception, and used to
		create dynamic models of the self and of the world''~\cite{runes62,DF13b}.
	\item Planning, namely the ability to make use of the Awareness models to compose
		a response to the changes being experienced.
\end{itemize}

A general scheme for robust system identity persistence is then given by the following three phases:
\begin{enumerate}
	\item Monitor the drifting of
		the $\Delta$ functions.
	\item Build models to understand how the drifting is impacting on one's
		system identity.
	\item Plan and enact corrective actions such that the system identity is not
		jeopardised.
\end{enumerate}
Phases 2. and 3. refer to the concept of behaviour---which is the subject of next subsection.

%----------------------------------%
\subsection{Behaviour}\label{s:BM:B}
%----------------------------------%
Behaviour is defined in~\cite{RWB43} as
``any change of an entity with respect to 
its surroundings\footnote{Here and in what follows, when not
	explicitly mentioned otherwise, quotes are from~\cite{RWB43}.}''.
In the context of this paper behaviour is to be meant as \emph{any change
an entity enacts in order to sustain their system identity}. In other words,
behaviour is the response a system enacts in order to be resilient.
In the cited paper the authors discuss how the above mentioned response
may range from simple and predefined reflexes up to complex context-aware strategies. The following classes are identified:

\begin{enumerate}
\item Passive behaviour: the system is inert, namely unable to produce any
``output energy''.
\item Active, non-purposeful behaviour. Systems in this
class, albeit ``active'', do not have a ``specific final condition toward
which they strive''.
\item Purposeful, non-teleological (i.e., feedback-free)
behaviour. A typical example of systems exercising this type
of behaviour is given by servo-mechanisms.
\item Teleological, non-extrapolative behaviours
	are those typical of \emph{reactive\/} systems. A feedback channel provides
	those systems with ``signals from the goal''. Behaviour is then
	adjusted in order to get ``closer'' to the goal as it was perceived
	through the channel. Reactive systems function under the
	implicit hypothesis that the adjusted behaviours bring indeed
	the system closer to the goals.
\item Predictive behaviours are typical of \emph{proactive\/} systems,
namely systems that base their action upon a hypothesised future state
computed through some model. In~\cite{RWB43} predictive behaviours
are further classified
according to their ``order'', namely the amount of
context variables their models take into account. Thus a system 
tracking the speed of another system to anticipate its future position exhibits
first-order predictive behaviours, while one that considers, 
e.g., speed and flightpath, is second-order predictive. Systems 
constructing their models through the correlation
of two or more ``raw fact'' dimensions, possibly of different nature,
are called higher-order predictive systems.
\end{enumerate}

%A second dimension in the behavioural responses is given by the invidualistic vs. collective nature
%of those responses~\cite{AF83,SW09,Sou00,Gus12}.
The above model of individual behaviour may be naturally extended by considering
collective behaviours---namely the conjoint behaviours of multiple individual systems.
We distinguish three major classes of collective behaviour:
\begin{enumerate}
	\item Neutral social behaviour. This is the behaviour resultant from the
		collective action of individual, purposeful, non teleological behaviours.
		Each participant operates through simple
		reflexes, e.g., ``in case of danger get closer to the flock''.
		Lacking a ``signal from the goal'', the rationale of this class of collective behaviours
		lies in the benefits deriving by the sheer number of replicas available.
		Examples include defencive behaviour of a group of individuals from a predator
		and group predation.
	\item Individualistic social behaviour. This is the social behaviour of systems
		trying to benefit opportunistically in a regime of competition with other systems.
		Here participants make use of more complex behaviours that take into
		account the social context, namely the behaviours exercises by the
		other participants. It is worth noting how even simple ``systems'' such
		as bacteria may exercise this class of behaviour~\cite{SW09,ML12}.
	\item Cooperative or coopetitive social behaviours. These are social behaviours of systems
		able to establish mutualistic relationships (mutually satisfactory behaviours) and
		to consider proactively the future returns deriving from a loss in the present.
		Examples of behaviours in this class
		are, e.g., the symbiotic relationships described in~\cite{SDGB10b,4262665}.
\end{enumerate}

As a final remark we deem worth noting how
resilience and change tolerance are not absolute properties:
in fact they emerge from the match with the particular conditions being exerted by the current environment. This
means that it is not possible to come up with an ``all perfect'' solution able to withstand whatever
such condition. Nature's answer to this dilemma is given by
redundancy and diversity. Redundancy and diversity are in fact key defence
frontlines against turbulent and chaotic events
affecting catastrophically an (either digital or natural) ecosystem.
Multiple and diverse ``designs'' are confronted with events
that determine their fit. Collective behaviours increase the chance that not all the designs will be negatively affected.
In this sense we could say that ``resilience abhors a vacuum'',
for empty spaces---namely unemployed designs and missed diversity---may
potentially correspond to the
very solutions that would be able to respond optimally
to a catastrophic event.

%(e.g.; stg happened in the Jurassic (?) which made dinosaurs extinct;
%but the ecosystem SURVIVED filling in the missing roles.
%Survival abhors empty spaces, 'cause they could correspond to the ``right'' responses
%to a castrophic event
A treatise of collective behaviours is outside the scope of this paper.
Interested readers may refer to, e.g., \cite{AF83,SW09,Sou00,Gus12}.

%-------------------------------------------------------------------------%
\subsection{Fragility as a measure of assumptions dependence}\label{s:BM:M}
%-------------------------------------------------------------------------%
The type of behaviour exercised by a system constitutes---we deem---a second
important characteristic of that system with reference to its ability
to improve dynamically its system-environment fit.
This ``second coordinate'' of a system's fidelity to
systemic, operational, and environmental assumptions is meant to bring to the foreground
how dependant an open system actually is on its \textbf{system model}---namely, on its
prescribed assumptions and working conditions~\cite{DeDe02b,DeFl09}.
%In the framework of real-timeliness, we can observe how the current trend is
%moving from synchronous system model assumptions requiring immutable
%platform and environmental conditions to more asynchronous or hybrid
%models~\cite{CrFe99} where fluctuations and service interruptions are
%contemplated as natural events rather than unlikely exceptions~\cite{MuCa11}.

Classes of resilient behaviours allow us to assess qualitatively a system's ``fragility'' (conversely, robustness)
to the variability of its environmental and systemic conditions.
As an example, let us consider the case of
traditional electronic systems such as, e.g., the flight control system that was in use in the
maiden flight of the Ariane 5 rocket. A common trait in such
systems is that enacted behaviours are mostly
very simple (typically purposeful but non-teleological). While this enhances efficiency and results
in a lean and cost-effective design, one may observe that it also produces a strong dependence 
on prescribed environmental conditions. It was indeed a mismatch between the prescribed and
the experienced conditions that triggered the chain of events that resulted in the Ariane 5 failure~\cite{De10}.

%%%%%%%%%%%%%%%%%%%%%%%%%%%%%%%%%%%%%%%%%%%%%%%%%%%%%%%%%%
\section{Beyond both Elasticity and Resilience}\label{s:D}
%%%%%%%%%%%%%%%%%%%%%%%%%%%%%%%%%%%%%%%%%%%%%%%%%%%%%%%%%%
We have introduced two complementary models to reason about the fidelity
of open systems.
The two models are orthogonal, in the sense that they represent two independent ``snapshots'' of the system under consideration:
\begin{enumerate}
	\item The Algebraic model regards the system as a predefined, immutable entity. The system conditions may drift
		but the system exhibits no complex ``motion''---no 
		sophisticated active behaviours
		are foreseen in order to reduce the drift.
		%In the case of hard- and soft real-timeliness, the system can
		%measure its system-environment fit, but usually can not
		%change it.
	\item The behavioural model captures instead the dynamicity of the
system. Also in this case the system measures
		the system-environment fit, but the system may actively use this measure in order to optimise its quality and execution.
\end{enumerate}

Intuitively, the first model is backed by redundant resources dimensioned through worst-case analyses; events potentially able
to jeopardise quality are \emph{masked\/} out. The minimal non-functional activity translates in low overhead and simple design.
Embedded systems typically focus on this approach.

Conversely, the second model calls for complex abilities---among others awareness; reactive and proactive planning; quorum sensing~\cite{SW09};
collective strategy planning~\cite{AF83}, etc. Events jeopardising quality are \emph{tolerated\/} rather than masked; moreover,
complex analyses and
strategies are mandated by the overhead typically associated with non-functional behaviours. This notwithstanding,
said behaviours may be the only effective line of defence against the highly dynamic environments characterising
\emph{open\/} embedded systems (such as cyber-physical things~\cite{Lee:EECS-2008-8}) and, \emph{a fortiori},
future collective cyber-physical societies~\cite{Zhu10} and fractal social organisations~\cite{DF13c}.

Though orthogonal in their assumptions, the overheads associated
with the design approaches corresponding to the above two models
are not side-effect free (suffice it to consider the effect of
complex behaviours on
the worst-case analyses called for, e.g., by hard real-time systems).
Our tentative answer to this problem is given by a new general scheme revising
the one presented in Sect.~\ref{s:BM:R}. In the new scheme the systems perform as follows:

\begin{enumerate}
	\item Monitor the drifting of
		their $\Delta$ functions and possibly other context figures
		providing insight on the current environmental conditions.
	\item Build simple, low-overhead models of the turbulence and chaotic nature of
		their environments.
	\item While the current conditions and trend are deemed as ``unsafe'', repeat:
	\begin{enumerate}
		\item Build and maintain more complex reactive and proactive models to understand
			how the drifting is impacting on one's system identity.
		\item Plan and enact corrective behaviours choosing between the following two options:
		\begin{enumerate}
			\item Self-reconfiguration: the system reshapes itself by choosing new
				system structures and new designs best matching the new environmental
				conditions. Examples of self-reconfiguration strategies may be found,
				e.g., in~\cite{De10}.
			\item Establish social relationships with neighbouring systems.
				This may include for instance simple actions such as ``join collective system''
				or ``leave collective system'',
				opportunistic strategies such as
				``improve one's $\Delta$'s to the detriment of those of neighbouring systems'',
				or complex mutualistic relationships involving
				association, symbiosis, or mutual assistance. An example of said
				mutualistic relationships is described in~\cite{DF13b}.
		\end{enumerate}
	\item Measure the effectiveness of the attempted solutions, rank them with respect to
		past solutions, derive and persist conclusions,
	and update the reactive and proactive models accordingly.\label{i:c}
	\end{enumerate}
\end{enumerate}
 
As can be clearly seen from its structure, the above scheme distinguishes two conditions: one in which
system identity is not at stake, and correspondingly
complexity and overhead are kept to a minimum, and one when new conditions are emerging that
may result in identity failures---in which case the system switches to more complex behaviours.
A self-managed, dynamic trade-off between these two approach, we conjecture, may provide designers with a solution
reconciliating the benefits and costs of both options. We refer to future systems able to exercise said dynamic trade-offs
as to ``auto-resilient''---a concept first sketched in~\cite{DF13b}.

As a final remark, the machine learning step~\ref{i:c} in the above scheme implies that the more a system is subjected to threats
and challenging conditions, the more insight will be acquired on how to respond to new and possibly more threatening
situations. We conjecture that insight in this process
may provide the designers with guidelines for engineering
\emph{antifragile\/} cyber-physical systems~\cite{Taleb12}.

\section{Conclusions}\label{s:C}
We presented two orthogonal models for the synchrony and real-timeliness of open computer systems
such as modern electronic systems~\cite{MuCa11}, cyber-physical systems, and collective organisations thereof.
We discussed how each of the two models 
best-match certain operational conditions---the former, stability; the latter,
dynamicity and turbulence. Finally, we proposed a scheme able to self-optimise
system processing depending on the experienced environmental conditions.
As the scheme also includes a machine learning step potentially able
to enhance the ability of the system to adjust to adverse environmental
conditions we put forward
the conjecture that antifragile systems may correspond to
systems able to learn while enacting elastic and resilient strategies.
Future work will be devoted to simulating compliant systems with the support of self-adaptation frameworks
such as ACCADA~\cite{Gui2011185,GuD10+} and Transformer~\cite{GDF12,GDFH13}.
%The bibliography, done here without a bib file
%This is the old BibTeX style for use with llncs.cls
%\bibliographystyle{splncs}
\bibliographystyle{IEEEtran}

%Alternative bibliography styles:
%the following does the same as above except with alphabetic sorting
%\bibliographystyle{splncs_srt}
%the following is the current LNCS BibTex with alphabetic sorting
%If you want to use a different BibTex style include [oribibl] in the document class line
%\bibliography{/refs/thesis}

\begin{thebibliography}{10}
\providecommand{\url}[1]{#1}
\csname url@samestyle\endcsname
\providecommand{\newblock}{\relax}
\providecommand{\bibinfo}[2]{#2}
\providecommand{\BIBentrySTDinterwordspacing}{\spaceskip=0pt\relax}
\providecommand{\BIBentryALTinterwordstretchfactor}{4}
\providecommand{\BIBentryALTinterwordspacing}{\spaceskip=\fontdimen2\font plus
\BIBentryALTinterwordstretchfactor\fontdimen3\font minus
  \fontdimen4\font\relax}
\providecommand{\BIBforeignlanguage}[2]{{%
\expandafter\ifx\csname l@#1\endcsname\relax
\typeout{** WARNING: IEEEtran.bst: No hyphenation pattern has been}%
\typeout{** loaded for the language `#1'. Using the pattern for}%
\typeout{** the default language instead.}%
\else
\language=\csname l@#1\endcsname
\fi
#2}}
\providecommand{\BIBdecl}{\relax}
\BIBdecl

\bibitem{Hey98}
\BIBentryALTinterwordspacing
F.~Heylighen, ``Basic concepts of the systems approach,'' in \emph{Principia
  Cybernetica Web}, F.~Heylighen, C.~Joslyn, and V.~Turchin, Eds.\hskip 1em
  plus 0.5em minus 0.4em\relax Principia Cybernetica, Brussels, 1998. [Online].
  Available: \url{http://pespmc1.vub.ac.be/SYSAPPR.html}
\BIBentrySTDinterwordspacing

\bibitem{MuCa11}
S.~Munaga and F.~Catthoor, ``Systematic design principles for cost-effective
  hard constraint management in dynamic nonlinear systems,''
  \emph{International Journal of Adaptive, Resilient and Autonomic Systems},
  vol.~2, no.~1, 2011.

\bibitem{Lee:EECS-2008-8}
\BIBentryALTinterwordspacing
E.~A. Lee, ``Cyber physical systems: Design challenges,'' EECS Department,
  University of California, Berkeley, Tech. Rep. UCB/EECS-2008-8, Jan 2008.
  [Online]. Available:
  \url{http://www.eecs.berkeley.edu/Pubs/TechRpts/2008/EECS-2008-8.html}
\BIBentrySTDinterwordspacing

\bibitem{AF83}
W.~Astley and C.~J. Fombrun, ``Collective strategy: Social ecology of
  organizational environments,'' \emph{The Academy of Management Review},
  vol.~8, pp. 576--587, October 1983.

\bibitem{Zhu10}
H.~Zhuge, ``Cyber physical society,'' in \emph{Semantics Knowledge and Grid
  (SKG), 2010 Sixth International Conference on}, Nov. 2010, pp. 1--8.

\bibitem{SDGB10b}
H.~Sun, V.~De~Florio, N.~Gui, and C.~Blondia, ``The missing ones: Key
  ingredients towards effective ambient assisted living systems,''
  \emph{Journal of Ambient Intelligence and Smart Environments}, vol.~2, no.~2,
  April 2010.

\bibitem{DeBl10}
V.~De~Florio and C.~Blondia, ``Service-oriented communities: Visions and
  contributions towards social organizations,'' in \emph{On the Move to
  Meaningful Internet Systems: OTM 2010 Workshops}, ser. Lecture Notes in
  Computer Science, R.~Meersman, T.~Dillon, and P.~Herrero, Eds.\hskip 1em plus
  0.5em minus 0.4em\relax Springer Berlin / Heidelberg, 2010, vol. 6428, pp.
  319--328.

\bibitem{Latour06}
B.~Latour, ``On actor-network theory. a few clarifications plus more than a few
  complications,'' \emph{Soziale Welt}, vol.~47, pp. 369--381, 1996.

\bibitem{DF13c}
V.~{De~Florio}, M.~Bakhouya, A.~Coronato, and G.~{Di Marzo Serugendo}, ``Models
  and concepts for socio-technical complex systems: Towards fractal social
  organizations,'' \emph{Systems Research and Behavioral Science}, vol.~30,
  no.~6, 2013.

\bibitem{DB07a}
V.~De~Florio and C.~Blondia, ``Reflective and refractive variables: A model for
  effective and maintainable adaptive-and-dependable software,'' in \emph{Proc.
  of the 33rd EUROMICRO Conference on Software Engineering and Advanced
  Applications (SEAA 2007)}, L{\"u}beck, Germany, August 2007.

\bibitem{Kanai12}
\BIBentryALTinterwordspacing
R.~Kanai and N.~Tsuchiya, ``Qualia,'' \emph{Current Biology}, vol.~22, no.~10,
  pp. R392--R396, 2012. [Online]. Available:
  \url{http://www.emotion.caltech.edu/~naotsu/Naotsugu_Tsuchiyas_homepage/papers_for_download_files/Kanai2012Current\%20Biology.pdf}
\BIBentrySTDinterwordspacing

\bibitem{DeFl09}
V.~De~Florio, \emph{Application-layer Fault-Tolerance Protocols}.\hskip 1em
  plus 0.5em minus 0.4em\relax IGI Global, Hershey, PA, 2009.

\bibitem{CrFe99}
F.~Cristian and C.~Fetzer, ``The timed asynchronous distributed system model,''
  \emph{IEEE Trans. on Parallel and Distributed Systems}, vol.~10, no.~6, pp.
  642--657, June 1999.

\bibitem{Taleb12}
N.~N. Taleb, \emph{Antifragile: Things That Gain from Disorder}.\hskip 1em plus
  0.5em minus 0.4em\relax Random House Publishing Group, 2012.

\bibitem{DF13b}
V.~{De Florio}, ``Preliminary contributions towards auto-resilience,'' in
  \emph{Proceedings of the 5th International Workshop on Software Engineering
  for Resilient Systems (SERENE 2013), Lecture Notes in Computer Science
  8166}.\hskip 1em plus 0.5em minus 0.4em\relax Kiev, Ukraine: Springer,
  October 2013, pp. 141--155.

\bibitem{DF12a}
V.~De~Florio, ``On the role of perception and apperception in ubiquitous and
  pervasive environments,'' in \emph{Proceedings of the 3rd Workshop on Service
  Discovery and Composition in Ubiquitous and Pervasive Environments
  (SUPE'12)}, August 2012.

\bibitem{Bou56}
K.~Boulding, ``General systems theory---the skeleton of science,''
  \emph{Management Science}, vol.~2, no.~3, April 1956.

\bibitem{leibniz2006shorter}
\BIBentryALTinterwordspacing
G.~Leibniz and L.~Strickland, \emph{The shorter Leibniz texts: a collection of
  new translations}, ser. Continuum impacts.\hskip 1em plus 0.5em minus
  0.4em\relax Continuum, 2006. [Online]. Available:
  \url{http://books.google.be/books?id=oFoCY3xJ8nkC}
\BIBentrySTDinterwordspacing

\bibitem{De10}
\BIBentryALTinterwordspacing
V.~De~Florio, ``Software assumptions failure tolerance: Role, strategies, and
  visions,'' in \emph{Architecting Dependable Systems VII}, ser. Lecture Notes
  in Computer Science, A.~Casimiro, R.~de~Lemos, and C.~Gacek, Eds.\hskip 1em
  plus 0.5em minus 0.4em\relax Springer Berlin / Heidelberg, 2010, vol. 6420,
  pp. 249--272, 10.1007/978-3-642-17245-8\_11. [Online]. Available:
  \url{http://dx.doi.org/10.1007/978-3-642-17245-8\_11}
\BIBentrySTDinterwordspacing

\bibitem{p3air}
R.~Charette, ``Electronic devices, airplanes and interference: Significant
  danger or not?'' January 2011, iEEE Spectrum blog ``Risk Factor'', available
  online through
  \url{http://spectrum.ieee.org/riskfactor/aerospace/aviation/electronic-devices-airplanes-and-interference-significant-danger-or-not}.

\bibitem{DF13a}
\BIBentryALTinterwordspacing
V.~{De~Florio}, ``On the constituent attributes of software and organisational
  resilience,'' \emph{Interdisciplinary Science Reviews}, vol.~38, no.~2, June
  2013. [Online]. Available:
  \url{http://www.ingentaconnect.com/content/maney/isr/2013/00000038/00000002/art00005}
\BIBentrySTDinterwordspacing

\bibitem{Sachs}
J.~Sachs, \emph{Aristotle's Physics: A Guided Study}, ser. Masterworks of
  Discovery.\hskip 1em plus 0.5em minus 0.4em\relax Rutgers University Press,
  1995.

\bibitem{DeAnimaLawsonTancred}
\BIBentryALTinterwordspacing
Aristotle and H.~Lawson-Tancred, \emph{De Anima (On the Soul)}, ser. Penguin
  classics.\hskip 1em plus 0.5em minus 0.4em\relax Penguin Books, 1986.
  [Online]. Available: \url{http://books.google.be/books?id=KsoFvv8vdM8C}
\BIBentrySTDinterwordspacing

\bibitem{runes62}
D.~D. Runes, Ed., \emph{Dictionary of Philosophy}.\hskip 1em plus 0.5em minus
  0.4em\relax Philosophical Library, 1962.

\bibitem{RWB43}
\BIBentryALTinterwordspacing
A.~Rosenblueth, N.~Wiener, and J.~Bigelow, ``Behavior, purpose and teleology,''
  \emph{Philosophy of Science}, vol.~10, no.~1, pp. 18--24, 1943. [Online].
  Available: \url{http://www.journals.uchicago.edu/doi/abs/10.1086/286788}
\BIBentrySTDinterwordspacing

\bibitem{SW09}
D.~Schultz, P.~G. Wolynes, E.~Ben~Jacob, and J.~N. Onuchic, ``Deciding fate in
  adverse times: Sporulation and competence in bacillus subtilis,'' \emph{Proc.
  Nat.l Acad. Sci.}, vol. 106, pp. 21\,027--21\,034, 2009.

\bibitem{ML12}
\BIBentryALTinterwordspacing
B.~McCallum and S.~Lam, ``Bacteria use `chemical twitter' and `prisoner's
  dilemma' to make decisions,'' \emph{Epoch Times}, March 2012. [Online].
  Available:
  \url{http://www.theepochtimes.com/n2/science/bacteria-use-chemical-twitter-and-prisoner-s-dilemma-to-make-decisions-211625.html}
\BIBentrySTDinterwordspacing

\bibitem{4262665}
H.~Sun, V.~De~Florio, N.~Gui, and C.~Blondia, ``Participant: A new concept for
  optimally assisting the elder people,'' in \emph{Computer-Based Medical
  Systems, 2007. CBMS '07. Twentieth IEEE International Symposium on}, june
  2007, pp. 295 --300.

\bibitem{Sou00}
P.~Sousa, N.~Silva, T.~Heikkila, M.~Kallingbaum, and P.~Valcknears, ``Aspects
  of co-operation in distributed manufacturing systems,'' \emph{Studies in
  Informatics and Control Journal}, vol.~9, no.~2, pp. 89--110, 2000.

\bibitem{Gus12}
K.~Guseva, ``Formation and cooperative behavior of protein complexes on the
  cell membrane,'' Ph.D. dissertation, Institute of Complex Systems and
  Mathematical Biology of the University of {A}berdeen, 2012.

\bibitem{DeDe02b}
V.~De~Florio and G.~Deconinck, ``On some key requirements of mobile application
  software,'' in \emph{Proc. of the 9th Annual IEEE International Conference
  and Workshop on the Engineering of Computer Based Systems (ECBS)}.\hskip 1em
  plus 0.5em minus 0.4em\relax Lund, Sweden: IEEE Comp. Soc. Press, April 2002.

\bibitem{Gui2011185}
\BIBentryALTinterwordspacing
N.~Gui, V.~De~Florio, H.~Sun, and C.~Blondia, ``Toward architecture-based
  context-aware deployment and adaptation,'' \emph{Journal of Systems and
  Software}, vol.~84, no.~2, pp. 185--197, 2011. [Online]. Available:
  \url{http://www.sciencedirect.com/science/article/B6V0N-512K1M8-2/2/2aa895c4f0dfbb40505396cfed5f1ba3}
\BIBentrySTDinterwordspacing

\bibitem{GuD10+}
------, ``{ACCADA}: A framework for continuous context-aware deployment and
  adaptation,'' in \emph{Proceedings of the 11th International Symposium on
  Stabilization, Safety, and Security of Distributed Systems, (SSS 2009)}, ser.
  Lecture Notes in Computer Science, R.~Guerraoui and F.~Petit, Eds., vol.
  5873.\hskip 1em plus 0.5em minus 0.4em\relax Lyon, France: Springer, November
  2009, pp. 325--340.

\bibitem{GDF12}
\BIBentryALTinterwordspacing
N.~Gui and V.~{De Florio}, ``Towards meta-adaptation support with reusable and
  composable adaptation components,'' in \emph{Proceedings of the sixth IEEE
  International Conference on Self-Adaptive and Self-Organizing Systems (SASO
  2012)}.\hskip 1em plus 0.5em minus 0.4em\relax IEEE, 2012. 

\bibitem{GDFH13}
\BIBentryALTinterwordspacing
N.~Gui, V.~{De~Florio}, and T.~Holvoet, ``Transformer: an adaptation framework
  with contextual adaptation behavior composition support,'' \emph{Software:
  Practice \& Experience}, 2012. 

\end{thebibliography}
%\input edcc14abr2.bbl
%\begin{thebibliography}{1}
%add each reference in here like this:
%\bibitem[RE1]{reference1}
%\end{thebibliography}


\end{document}